\documentclass[journal,twoside]{IEEEtran}
\usepackage{amsmath,amsfonts}
\usepackage{algorithmic}
\usepackage{array}
\usepackage{bm}
\usepackage[caption=false,font=normalsize,labelfont=sf,textfont=sf]{subfig}
\usepackage{textcomp}
\usepackage{stfloats}
\usepackage{verbatim}
\usepackage{graphicx}
\usepackage{cite} 
\usepackage{ifthen}  
\usepackage{multicol} 
\usepackage{dsfont} 
\usepackage{amssymb}
\usepackage{color} 
\usepackage{soul} 
\usepackage{booktabs}
\usepackage{ctable} 
\usepackage{balance}
\usepackage{tabularx} 
\usepackage{url}
\usepackage{hyperref}

\makeatletter

\newcolumntype{'}{!{\vrule width 1pt}}
\makeatother 
\def\BibTeX{{\rm B\kern-.05em{\sc i\kern-.025em b}\kern-.08em
    T\kern-.1667em\lower.7ex\hbox{E}\kern-.125emX}}
\begin{document}
\title{Emerging NGSO Constellations: Spectral Coexistence with GSO Satellite Communication Systems}
\author{Flor~Ortiz,~\IEEEmembership{Member,~IEEE,}
Eva~Lagunas,~\IEEEmembership{Senior Member,~IEEE,}
Almoatssimbillah Saifaldawla, ~\IEEEmembership{Graduate Student Member,~IEEE,}
Mahdis Jalali, ~\IEEEmembership{Graduate Student Member,~IEEE,}
Luis~Emiliani,~\IEEEmembership{Senior Member,~IEEE,} Symeon~Chatzinotas,~\IEEEmembership{Fellow~IEEE} }

\markboth{IEEE Communications Magazine}{F. Ortiz \MakeLowercase{\textit{(et al.)}: Emerging NGSO Constellations: Spectral Coexistence with GSO Satellite Communication Systems}}
\maketitle
\begin{abstract}
Global communications have undergone a paradigm shift with the rapid expansion of low-earth orbit (LEO) satellite constellations, offering a new space era of reduced latency and ubiquitous, high-speed broadband internet access. However, the fast developments in LEO orbits pose significant challenges, particularly the coexistence with geostationary earth orbit (GEO) satellite systems. This article presents an overview of the regulatory aspects that cover the spectrum sharing in the bands allocated to the Fixed Satellite Service between geostationary networks (GSO) and non-geostationary systems (NGSO), as well as the main interference mitigation techniques for their coexistence. Our work highlights the increased potential for inter-system interference. It explores the regulatory landscape following the World Radio Conference (WRC-23). We discuss the different interference management strategies proposed for the GSO-NGSO spectral coexistence, including on-board and ground-based approaches and more advanced mitigation techniques based on beamforming. Moving onto operational aspects related to the sharing of spectrum, we introduce recent work on interference detection, identification, and mitigation and provide our vision of the emerging role of artificial intelligence (AI) in the aforementioned tasks. 
\end{abstract}
\section{Introduction}
\label{sec:intro}
\IEEEPARstart{T}{he} recent developments in the deployment of Low Earth Orbit (LEO) satellite constellations mark a revolutionary change in space-based communications, high-throughput broadband internet access at reduced latency everywhere on the Globe. Non-geostationary systems (NGSO) are allocated to the same frequency bands as the geostationary (GSO)systems and the rapid emergence of NGSO mega constellations comes with an increase in the potential of interference between systems, both NGSO and GSO. Primarily, the concern lies with potential interference, which can undermine the reliability and quality of the communication services provided by satellite systems. This issue has attracted the attention of industry stakeholders, regulators, and academic researchers, resulting in a complex dialogue about the balance between innovation and the integrity of established satellite networks \cite{radiocommunication2018functional}.

Historically, GSO networks have been the mainstay of space-based communications, occupying a unique orbit that allows stable and continuous coverage of specific terrestrial regions. The emergence of NGSO systems, characterized by lower orbital altitude and low latency, introduces a dynamic environment where the risk of interference between GSO and NGSO systems increases significantly. Such interference threatens the quality of service and raises concerns about spectrum allocation and orbital debris \cite{ITU_2023,ravishankar2021next, 9977866}.

Integrating NGSO System into the current satellite spectrum framework poses a complex challenge that requires collaboration, innovation, and foresight. Now that we are in a new era of space communications, it is imperative to navigate these complexities to ensure the harmonious and sustainable coexistence of NGSO and NGSO systems. The fast deployment of mega-constellations by various companies underscores the need to re-evaluate existing regulations and develop new strategies to mitigate interference risks \cite{henri2024itu}.

The International Telecommunication Union (ITU) plays a key role in this area, mainly through its World Radiocommunication Conference (WRCs). The most recent conference, WRC-23, held in Dubai from 20 November to 15 December 2023, addressed several critical issues related to satellite communications \cite{ITU_2023b}. A key resolution of the preceding Plenipotentiary Conference (PP-22), "Sustainability of radio-frequency spectrum and orbital resources associated with satellites used by space services," called for urgent studies on the increasing use of radio-frequency spectrum and orbital resources in NGSO, such as those used by LEO constellations. The need for equitable spectrum access and rational use of orbital and spectrum resources was stressed for GSO and NGSO, in line with Article 44 of the ITU Constitution.


This article discusses the challenges posed by interference between GSO and NGSO systems. We will explore the current regulatory landscape, detailing the evolution of policies and regulations governing space-based communications. Next, we extend the discussion to the various technical approaches employed to avoid or minimize potential interference, highlighting traditional methods (e.g. exclusion angles, satellite tilting) and more advanced multi-antenna techniques (e.g. beamforming design). We revise potential interference detection and identification strategies, which are becoming increasingly popular among stakeholders and regulators, to verify and confirm that the radio regulations are being followed. Finally, we discuss the role of artificial intelligence (AI) in the aforementioned tasks, highlighting its suitability for such a time-varying and fast-changing wireless environment. Overall, the article aims to provide a global perspective on the efforts to balance the rapid advancement of NGSO systems and mega-constellations with the imperative to protect and maintain the services offered by other satellite systems and planned and in-operation communication systems.

\section{Spectrum Regulatory Landscape}
The evolution of satellite communications regulation has been driven by the increasing demand for -and complexity- of space services. The ITU Radio Regulations Treaty is crucial in managing these demands by allocating radio frequencies and providing a framework for coordinating orbit-spectrum resources. This framework is necessary to ensure reliable operation in space, given the finite nature of the radio spectrum and orbits around the Earth \cite{davies2021approaches}, as well as balancing the multi-stakeholder economic interests.

To minimize harmful interference and ensure operability, administrations apply to the ITU for orbital slots for GSO satellites and constellation parameters for NGSO satellites. The ITU then reviews these applications against necessary conditions for the proposed frequency range of operation and either accepts the application or returns it with comments for further adjustments. This critical review process is essential to optimize spectrum and orbital resources and prevent frequency warehousing. Following a successful review, the next step involves coordination based on the application's specifics and the proposed frequency ranges. Adherence to the Radio Regulation articles relevant to the requested bands is crucial \cite{millwood2023urgent}.

The successful operation of a satellite communication system within the Fixed Satellite Service (FSS) frequencies necessitates a comprehensive progression through several phases, from the system's conceptualization to its final launch and arrival at the intended orbit. From a regulatory standpoint, this process entails a series of preliminary steps aimed at informing all relevant parties of the planned new system. This is achieved through either an Advanced Publication (API) or a Coordination Request (CRC), depending on the system type and the frequencies required. Following this, the process continues with inter-system coordination to ensure compatibility with existing or planned systems operating within the same frequency bands, addressing potential conflicts with the network likely to be affected.

Subsequently, the filing enters the notification and bring into use procedures, in which the filing parameters may be adjusted to reflect the outcomes of the coordination discussions. The final stage of the regulatory process is the registration of the frequency assignments into the Master
International Frequency Register (MIFR). In the case of NGSO networks,  depending on the frequency band of operations, the ITU-R also requires operators to meet spacecraft deployment targets as part of a regulatory milestone procedure. Interested readers are encouraged to delve into the materials available from the ITU-R, particularly those from their space workshops, for a more in-depth understanding of these regulatory procedures \cite{henri2024itu}.

The Radio Regulations have not universally prioritized GSO over the NGSO system; this priority exists specifically within certain portions of the Fixed Satellite Service (FSS). The protection for GSO networks is principally maintained through a regulatory mechanism involving the Equivalent Power Flux Density (EPFD), which establishes limits on the emissions from NGSO constellations. This mechanism ensures that emissions from NGSO systems do not cause harmful interference to the GSO system and its terminals.

It's crucial to differentiate that the EPFD regulations and international coordination efforts address different aspects and portions of the spectrum band. As the demand for frequencies and orbital resources intensifies, there's a growing dialogue among international stakeholders to ensure equitable access and the sustainability of space operations. Specifically, some administrations and NGSO system operators have raised discussions on the need to reassess the EPFD limits as delineated in the Radio Regulations, Article 22. 

The ongoing debate highlights differing viewpoints among stakeholders. The discussions are centred on finding a balanced approach that facilitates the success of planned satellite constellations while maintaining the integrity of existing systems. This evolving dialogue underscores the dynamic nature of space regulation and the importance of fostering an adaptable and inclusive regulatory environment \cite{pritchard2023wrc}. Current discussions revolve around three points:

\begin{itemize}
    \item \textbf{Maintaining the current EPFD levels:} Proponents of retaining the existing EPFD limits emphasize their long-standing role in providing stable and continuous coverage for communications. This view is supported by the need to maintain the reliability and quality of satellite services.
    \item \textbf{In favour of re-evaluating EPFD levels:} Proponents of re-evaluating the EPFD limits highlight these satellites' innovative and expanded capabilities, including higher bandwidth and lower latency communications. They advocate for regulatory adaptations to accommodate the burgeoning LEO sector, emphasizing the need for equitable access to spectrum and orbital resources for these new technologies.
    \item \textbf{Developing methods for calculating and verifying,  the aggregate EPFD produced by multiple systems:} The ITU-R has been tasked by the WRC-23 with developing recommendations to a) provide a methodology for calculating the aggregate co-frequency EPFD produced by non-GSO FSS systems and accurately modelling non-GSO FSS operations, and b) a suitable methodology to adapt the operation of co-frequency non-GSO FSS systems to ensure that the aggregate power levels are met.
\end{itemize}

\section{Interference between GSO and NGSO systems}
Managing interference between GSO and NGSO systems presents complex challenges that necessitate understanding the specific characteristics and operational environments of these satellite systems. This section outlines the different interference scenarios and mitigation strategies.

\subsubsection{GSO-NGSO Interference}
Interference from GSO satellites to NGSO systems is primarily a concern in the uplink scenario, where ground stations transmitting to GSO satellites can produce significant power flux density (PFD) at NGSO satellites' antennas within the same frequency bands. This situation underscores the critical importance of the ITU-R regulatory framework and conducting frequency coordination following the ITU-R RR procedures.

Advanced beam-steering techniques by GSO satellites and adaptive power control by ground stations can significantly reduce the risk of interference with NGSO systems for frequency bands where coordination is required.

\subsubsection{NGSO-to-GSO Interference}
NGSO systems, with their dynamic orbits and coverage patterns, can interfere with GSO satellites due to their shared use of the FSS frequency range. To mitigate this, NGSO systems can utilize adaptive frequency allocation, beamforming, and power control strategies. These methods adjust frequency assignments and power levels based on the proximity and trajectory of GSO satellites, ensuring minimal interference. 

Significantly, the motion of NGSO satellites and the variation in elevation angles are not inherently problematic. The critical factor is the emission level, which must be managed carefully to avoid exceeding the interference-to-noise ratio (I/N) thresholds as NGSO satellites transit across the coverage areas of GSO ground stations.

\subsubsection{NGSO-to-NGSO Interference}
The risk of interference among NGSO satellites, whether co-channel or inter-system, arises from the density of satellites in orbit and their shared frequency bands. Effective mitigation requires coordinated frequency assignments and sophisticated interference management techniques. These include interference cancellation algorithms and adaptive modulation, which help maintain co-channel interference within acceptable levels and avoid spectral congestion.

The regulatory framework emphasizes coordination between systems rather than managing intra-system interference, which is considered a design parameter for each constellation. 

\section{On-Board vs. On-Ground Interference Management}
Interference management strategies, whether implemented on-board the satellite or from ground control centres, are crucial for the seamless operation of satellite networks. Each approach offers distinct advantages and addresses different aspects of interference management.

\subsubsection{On-board interference management}

On-board interference management focuses primarily on the uplink path, employing advanced algorithms and hardware to detect and mitigate interference directly from the satellite e.g. nulling in determined directions. This enables immediate response to interference events, particularly useful for managing emissions from inter-satellite links (ISLs) or ground stations within the satellite's network. Satellites' ability to autonomously adjust their operations to mitigate interference at its source or minimize impact is a significant advantage.

However, the capabilities of a satellite to mitigate interference originating from external sources, such as another operator's ground station (operator B), is limited. In such cases, the primary role of onboard systems is to detect interference and, where possible, adjust the satellite's operations to minimize impact.

\subsubsection{On-ground interference management}

On-ground interference management involves centralized control and coordination from ground-based facilities, enabling a comprehensive view of network operations and interference scenarios. This approach allows extensive computational analysis, historical data review, and predictive modelling to manage and mitigate interference.

Integrated space-ground systems represent a hybrid approach, combining onboard detection with ground-based command and control capabilities. This enables the dynamic allocation of spectrum resources and adjustment of satellite transmission patterns to nullify or reduce interference sources. Such systems facilitate direct communication with other operators to coordinate mitigation efforts, whether through operational parameter adjustments or central regulatory intervention.

The predominant approach in current operations involves a combination of on-board detection and ground-based control, emphasizing the need for cooperative efforts across different satellite operators to maintain interference within negotiated and regulated thresholds. This cooperative model reflects an advanced understanding of space operations, where strategic spectrum allocation and operational flexibility are key to mitigating interference across the increasingly crowded orbital environment.

\section{Interference Mitigation Techniques}

\begin{figure*}[!t]
  \centering
  \subfloat[Exclusion Angles]{\includegraphics[width=0.5\textwidth]{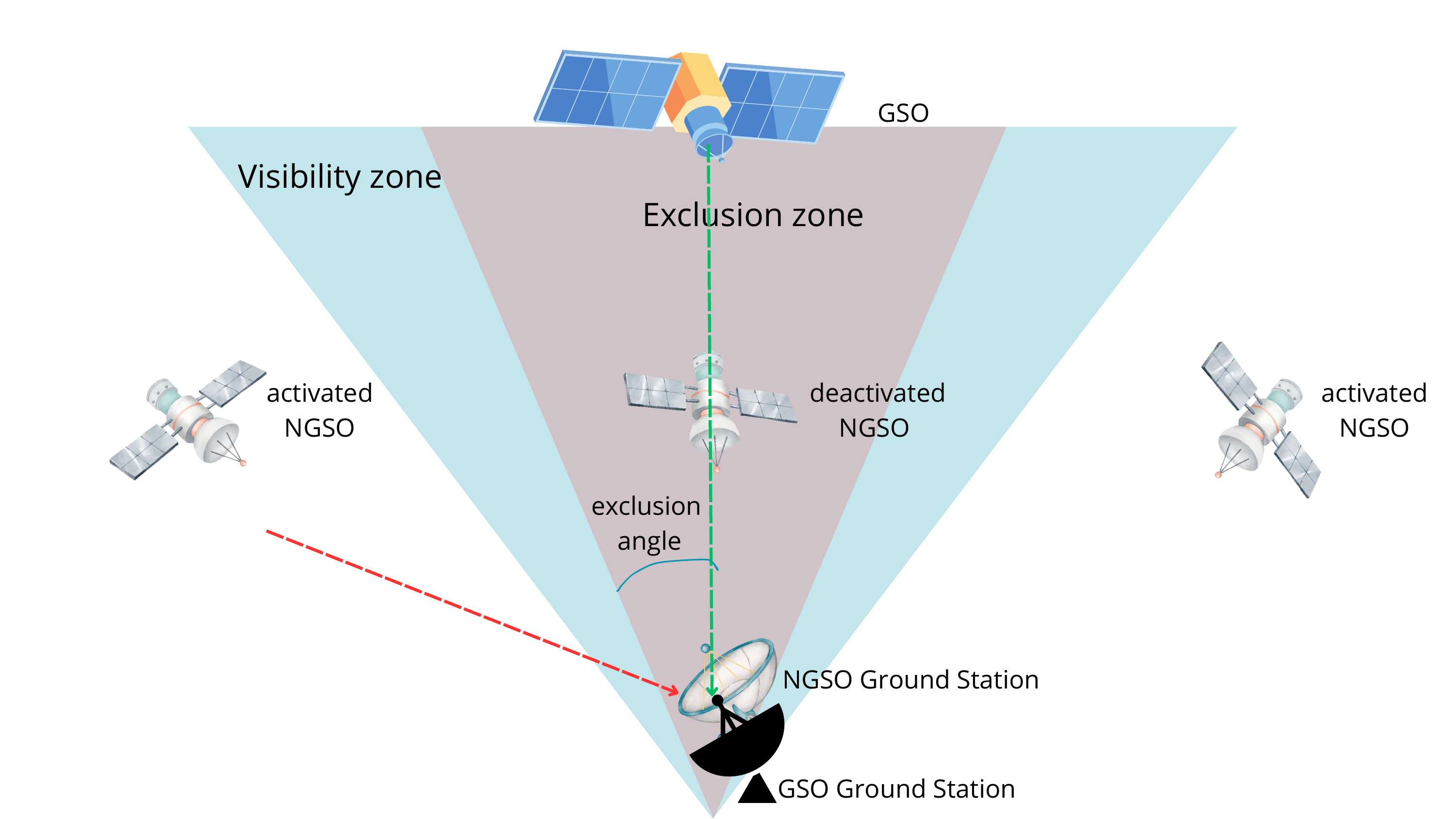}\label{fig:sub1}}
  \hfill
  \subfloat[Dynamic Power Adjustment]{\includegraphics[width=0.5\textwidth]{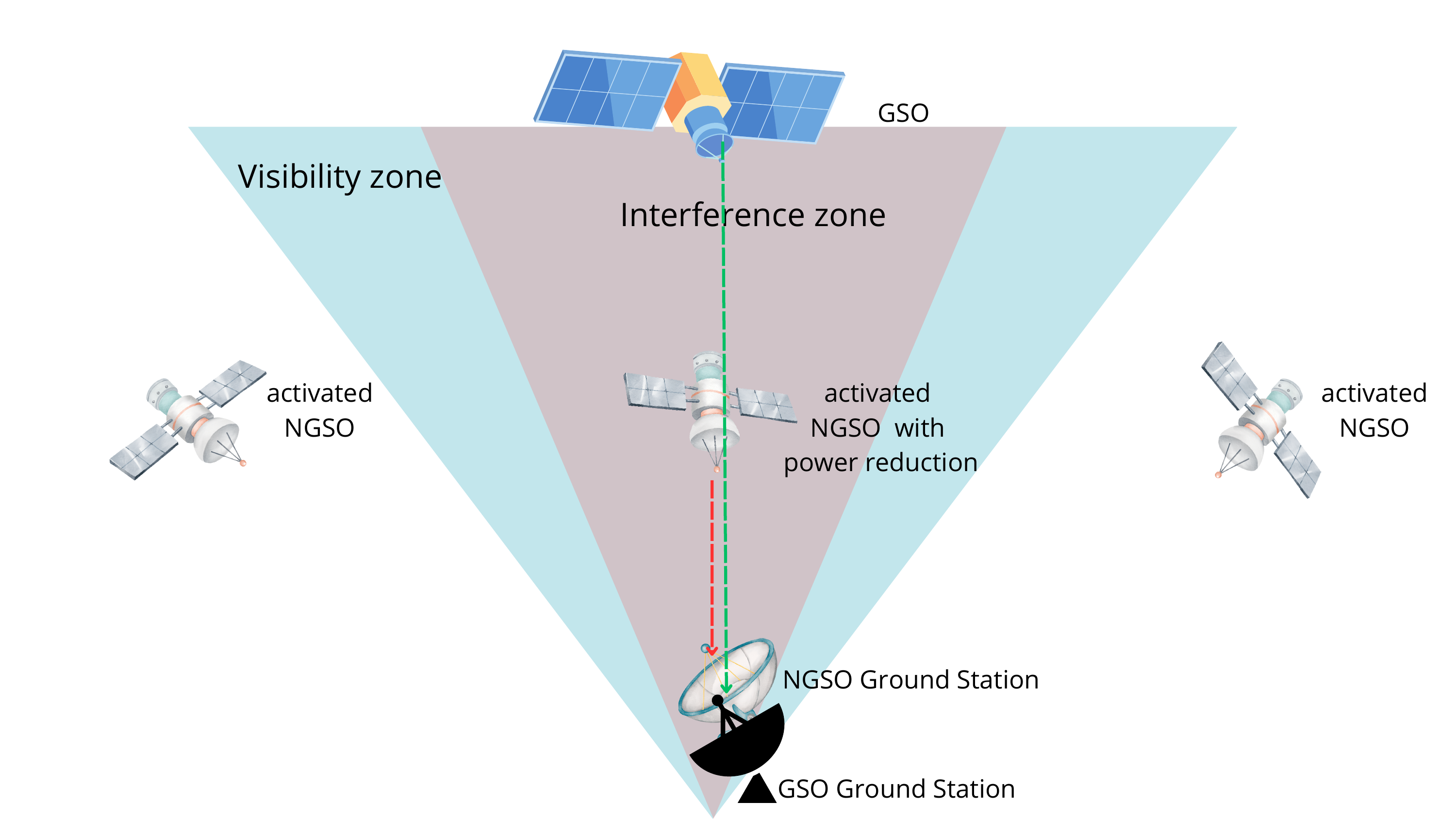}\label{fig:sub2}}
  \\
  \subfloat[Antenna Orientation Adjustment]{\includegraphics[width=0.5\textwidth]{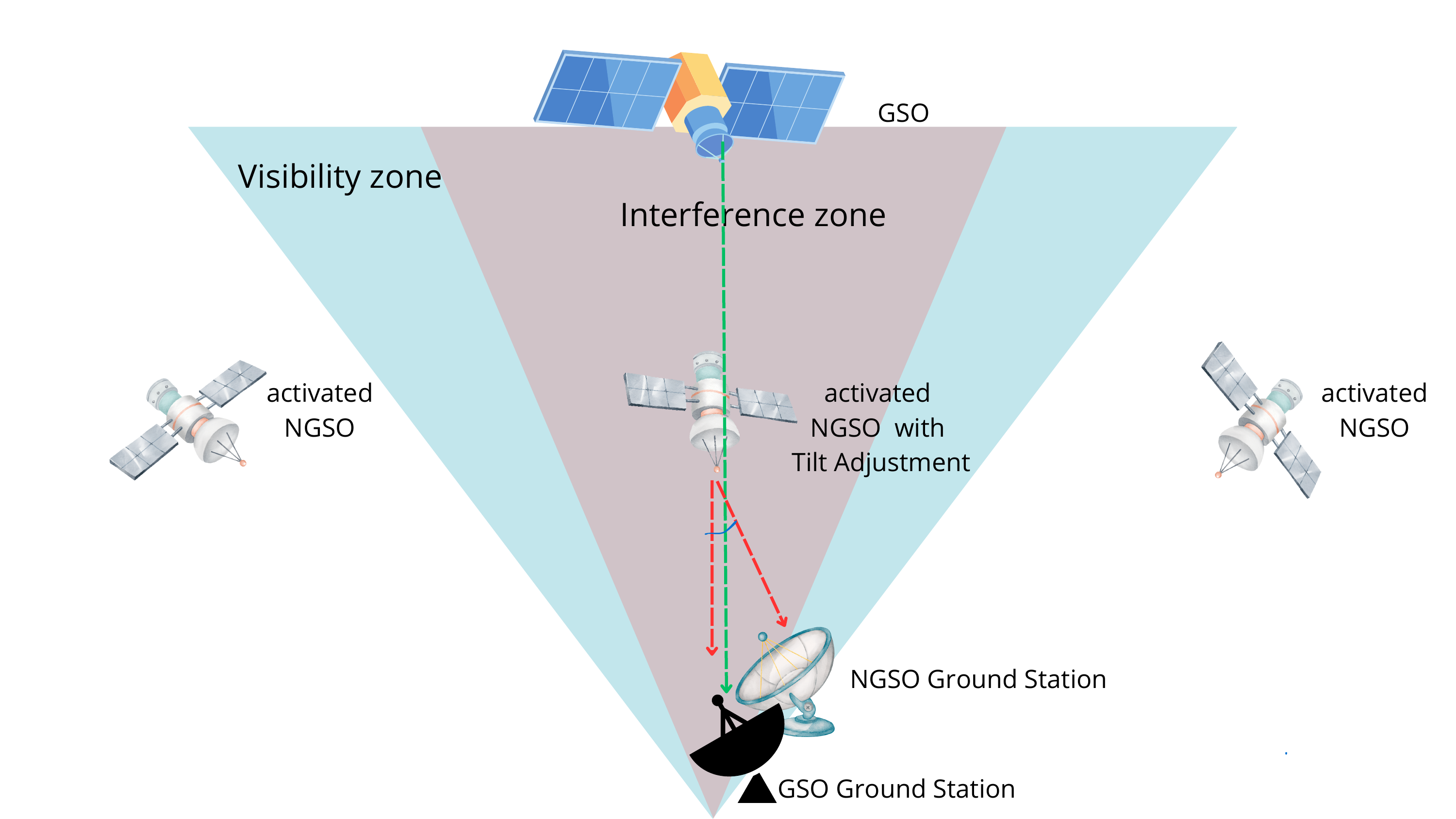}\label{fig:sub3}}
  \hfill
  \subfloat[Adaptive Beamforming]{\includegraphics[width=0.5\textwidth]{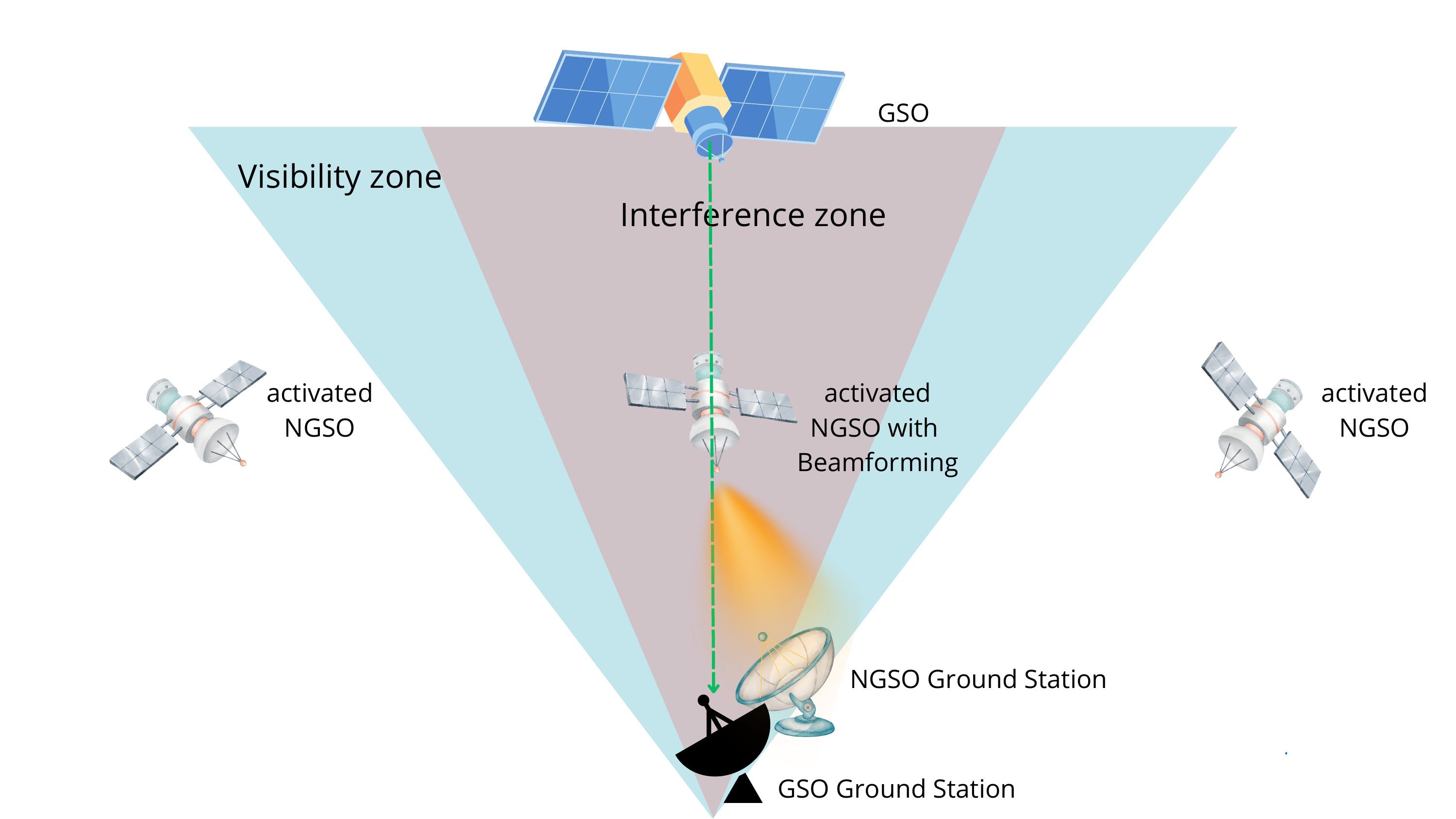}\label{fig:sub4}}
  \caption{Interference Mitigation Techniques}
  \label{fig:Mitigation_techniques}
\end{figure*}

Effective interference mitigation measures are crucial for the harmonious coexistence of NGSO systems with GSO systems, particularly as the number of NGSO satellites increases. Various techniques are employed, each varying in complexity and effectiveness. These techniques are illustrated in Fig. \ref{fig:Mitigation_techniques} and discussed below, referencing \cite{10209992, cottatellucci2006interference}:

\begin{itemize}
\item \textbf{Exclusion Zones:} This approach involves defining zones around GSO satellites where NGSO operations are restricted. When NGSO satellites approach these zones, they either modify their operational mode or adjust their transmission power to minimize interference. This strategy, while effective in reducing interference risks, may impact NGSO service continuity.
\item \textbf{Dynamic Power Adjustment:} NGSO satellites dynamically adjust their transmission power depending on their proximity to GSO satellites. By lowering transmission power when near GSO satellites, NGSO satellites reduce potential interference, maintaining service quality without interruptions.
\item \textbf{Antenna Orientation Adjustment:} This technique involves modifying the orientation of NGSO satellite antennas to minimize signal transmission towards GSO earth stations. Adjustments are made based on satellite location and the relative positions of GSO satellites, effectively reducing interference while allowing for overlap in service areas by neighboring NGSO satellites.
\item \textbf{Adaptive Beamforming:} Utilizing multiple antenna elements, NGSO satellites can precisely shape and steer their transmission beams away from GSO earth stations. Although complex, this technique is highly effective in minimizing interference and optimizing system capacity.
\end{itemize}

\begin{figure}[!t]
\centering
\includegraphics[width=0.5\textwidth]{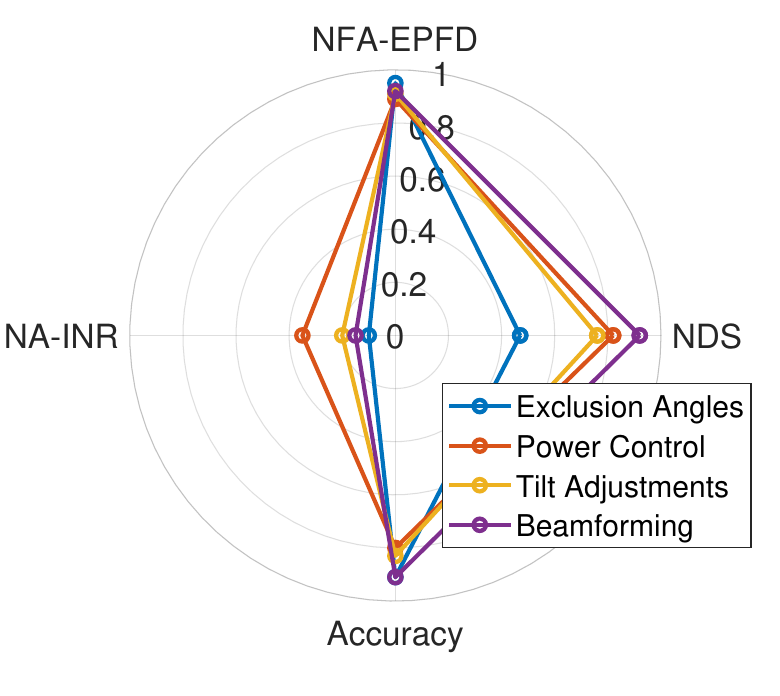}
\caption{Performance evaluation of different interference mitigation techniques between NGSO and GSO, highlighting trade-offs in effectiveness and user demand satisfaction.}
\label{fig:Performence_Mitigation_technique}
\end{figure}

Several metrics are key in evaluating the efficacy of interference mitigation techniques between NGSO and GSO satellite systems, each encapsulating distinct facets of interference management and system efficacy. These metrics include: (1) the \textit{Normalized Frequency of Achieving EPFD (NFA-EPFD)}, which gauges the proportion of time the EPFD is maintained at or below the threshold, normalized on a 0 to 1 scale, defined as $\text{NFA-EPFD} = \frac{\text{Time at or below threshold}}{\text{Total observation time}}$; (2) the \textit{NGSO User Demand Satisfaction (NDS)}, assessing the extent to which interference mitigation techniques fulfill NGSO user demands, normalized between 0 and 1, calculated by $\text{NDS} = \frac{\text{Satisfied User Demand}}{\text{Total User Demand}}$; (3) the \textit{Normalized Average Interference-to-Noise Ratio (NA-INR)}, evaluating the system-wide impact of interference relative to noise, normalized for benchmarking, expressed as $\text{NA-INR} = \frac{\text{Average INR}}{\text{Maximum Observed INR}}$; and (4) \textit{Mitigation Accuracy}, indicating the precision of a mitigation technique in reducing interference to agreed levels, though not defined by a specific equation in the original description. These metrics facilitate a comprehensive analysis in simulations featuring a constellation similar to the OneWeb LEO network, consisting of 1980 satellites across 36 orbits and a GSO satellite positioned over the equator at 20° longitude, providing insights into the techniques' performance in sustaining operational effectiveness and regulatory conformity within intricate satellite communication frameworks \cite{9977866}.

Fig. \ref{fig:Performence_Mitigation_technique} provides a comparative analysis of these mitigation techniques, with metrics such as the frequency of achieving EPFD at or below the maximum allowable level, user demand satisfaction (normalized), and the average interference-to-noise ratio. The evaluation considers a Walker-Star constellation as the NGSO system and a GSO satellite at 20° longitude, focusing on downlink scenarios.

Hybrid approaches that combine multiple techniques enhanced by AI are emerging as effective strategies for dynamic interference management. AI models, capable of anticipating interference scenarios based on satellite orbits and operational patterns, facilitate real-time adjustments to power levels, beam direction, and antenna orientation, enhancing mitigation effectiveness.

\section{Interference Detection and Identification}

Interference detection and identification are fundamental to effective interference mitigation, facilitating targeted responses to preserve service quality in densely populated orbital regions. This section discusses the roles of various stakeholders in interference mitigation and outlines detection techniques \cite{10293839}.

\subsection{Role in Interference Mitigation}
Accurate detection and identification enable satellite operators to apply specific mitigation strategies, crucial for regulatory compliance and service quality.

\subsubsection{Monitoring and Verification of Radio Regulation Compliance} For regulators, verifying adherence to ITU limits post-launch is challenging. Effective detection and identification mechanisms can help ensure equitable use of space resources and compliance with spectrum norms.

\subsubsection{Verification of Interference Levels} LEO operators benefit from methodologies that assess their satellites' emissions, aiding in compliance with regulations. Similarly, GEO operators need to identify interference sources and levels to protect their services.

\subsection{Detection Techniques}
\begin{figure}[!t]
\centering
\includegraphics[width=0.5\textwidth]{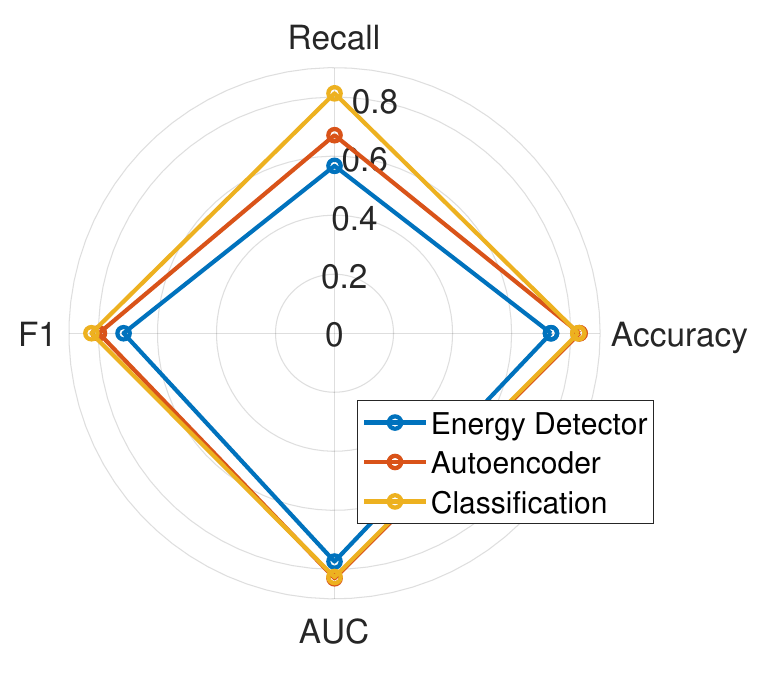}
\caption{Comparison of interference detection techniques in NGSO-GSO satellite communications, focusing on downlink scenarios.}
\label{fig:Performence_Detection_technique}
\end{figure}

Fig. \ref{fig:Performence_Detection_technique} compares traditional and AI-based detection methods across various performance metrics. Traditional methods like energy detection and spectral analysis offer simplicity but may falter in complex environments. Conversely, AI-driven techniques, such as autoencoders and classification models, demonstrate superior adaptability and learning capabilities, essential for dynamic satellite communication contexts.

AI-based methods stand out for their ability to discern between interfering and non-interfering signals accurately, with autoencoders showing promise for their flexibility and independence from labeled data sets. The comparison underscores AI's potential in improving interference detection and mitigation strategies, facilitating more reliable and efficient satellite communications.
\section{Discussion and Future Directions}
Integrating NGSO constellations into the current satellite communications framework has highlighted the complexity of managing interference between the established GSO systems and the high-mobility NGSO satellites. 

WRC-23 highlighted how the rapid advancement of new engineering solutions in satellite communications often outpaces the development of the international regulatory framework. This discrepancy presents significant challenges for administrations managing spectrum use for emerging technologies. It underscores the critical role of the ITU-R in ensuring seamless, interference-free operation and achieving global harmonization in satellite communications. Additionally, it's noted that several key issues related to satellite networks and systems are slated for further discussion at WRC-27. 

Notably, Agenda Item 1.6 for WRC-27 will focus on considering technical and regulatory measures for fixed-satellite service satellite networks/systems within the frequency bands 37.5-42.5 GHz (space-to-Earth), 42.5-43.5 GHz (Earth-to-space), 47.2-50.2 GHz (Earth-to-space), and 50.4-51.4 GHz (Earth-to-space). The goal is to ensure equitable access to these frequency bands, in accordance with existing resolutions.

Furthermore, Agenda Item 1.11 will address the technical and operational issues and regulatory provisions for space-to-space links among non-geostationary and geostationary satellites. These discussions will span frequency bands 1 518-1 544 MHz, 1 545-1 559 MHz, 1 610-1 645.5 MHz, 1 646.5-1 660 MHz, 1 670-1 675 MHz, and 2 483.5-2 500 MHz, which are allocated to the mobile-satellite service, in line with Resolution 249 (Rev.WRC-23). This forward-looking agenda highlights the ITU-R's commitment to addressing the evolving landscape of satellite communications, ensuring that regulatory frameworks keep pace with technological advancements to support smooth and harmonized global satellite operations.

Interference management, a significant concern for this integration, encompasses a variety of techniques such as off and exclusion angles, power control, tilt adjustments, and beamforming. The effectiveness of these techniques varies, with trade-offs between complexity and performance. 

The role of interference detection and identification has been highlighted as crucial, especially in light of regulatory compliance and effective mitigation. Traditional methods such as energy detection and spectral analysis have been juxtaposed with emerging AI techniques, including autoencoders and classification methods, revealing AI-based approaches' superior adaptability and efficiency.

The potential of AI to improve the effectiveness of interference mitigation is undeniable. The ability of AI to anticipate and preemptively address interference through real-time data analysis and predictive modeling offers a paradigm shift in how satellite networks are managed. Future research and development in this field could focus on refining AI algorithms for greater accuracy and adaptability in dynamic orbital environments.

On the other hand, recent missions by NASA, ESA, and other space agencies, along with initiatives by public and private organizations such as IRIS2 and Starlink, are pushing the boundaries of satellite technology and communication capabilities. These missions contribute to the advancement of space exploration and pose new challenges and opportunities in satellite communications, especially in managing increasingly saturated orbital trajectories and frequency spectrum.

Satellite communications are poised for changes driven by technological advances and growing global connectivity needs. Key trends and directions are as follows:
\begin{enumerate}
\item Increased global connectivity: expansion of satellite networks, especially through NGSO constellations, promises to bring high-speed Internet and communication services to remote and underserved regions, bridging the digital divide. NGSO operators want fast regulatory procedures to get operational licenses and be able to launch to space in timely manner. 

\item Sustainable space operations: As space becomes increasingly crowded, the focus will shift to sustainable operations, encompassing effective space debris management, responsible use of frequencies and environmental considerations. Complaints about satellite pollution and its impact on astronomy have already been raised, and how such a large number of flying devices would affect the amount of radiation received by the Earth are sensitive and emerging topics that we have never had to worry about.

\item Technological innovations: Continued advances in satellite technology, such as miniaturization, advanced propulsion systems and more efficient power sources, will enable more cost-effective and versatile satellite solutions.

\item Regulatory evolution: The regulatory framework governing satellite communications will need to continually evolve to address the dynamic nature of the sector, balancing innovation with the protection of existing services and the equitable allocation of resources. The mentality has to change. We are no longer dealing with a single massive and static satellite (as NGSO) but rather with dozens or hundreds of small satellites that move at high speed and are capable of steering the power dynamically to the coverage area on the Earth. Similarly, the small satellites have a much shorter lifespan and must be decommissioned.

\item Collaboration between government agencies, international organizations and private entities will be crucial in shaping the future of global satellite communications, ensuring coordination of efforts and knowledge sharing.
\end{enumerate}

\balance 
\bibliographystyle{IEEEbib}
\bibliography{refs_neurosat}

\begin{thebibliography}{10}

\bibitem{radiocommunication2018functional}
ITU Radiocommunication,
\newblock ``Functional description to be used in developing software tools for determining conformity of non-geostationary-satellite orbit fixed-satellite service systems or networks with limits contained in article 22 of the radio regulations [s/ol],''
\newblock 2018.

\bibitem{ITU_2023}
{ITU Hub},
\newblock ``Itu and space: Ensuring interference-free satellite orbits in leo and beyond,'' \url{https://www.itu.int/en/mediacentre/backgrounders/Pages/itu-and-space.aspx}, 2023,
\newblock Accessed: 2024-03-25.

\bibitem{ravishankar2021next}
Channasandra Ravishankar, Rajeev Gopal, Nassir BenAmmar, Gaguk Zakaria, and Xiaoling Huang,
\newblock ``Next-generation global satellite system with mega-constellations,''
\newblock {\em International journal of satellite communications and networking}, vol. 39, no. 1, pp. 6--28, 2021.

\bibitem{9977866}
Mahdis Jalali, Flor~G. Ortiz-Gomez, Eva Lagunas, Steven Kisseleff, Luis Emiliani, and Symeon Chatzinotas,
\newblock ``Radio regulation compliance of {NGSO} constellations' interference towards {GSO} ground stations,''
\newblock in {\em 2022 IEEE 33rd Annual International Symposium on Personal, Indoor and Mobile Radio Communications (PIMRC)}, 2022, pp. 1425--1430.

\bibitem{henri2024itu}
Yvon HENRI and Attila MATAS,
\newblock ``Itu regulatory regime related to non-gso satellite systems,''
\newblock {\em Space Law: Legal Framework for Space Activities}, p. 155, 2024.

\bibitem{ITU_2023b}
{ITU},
\newblock ``Itu-r preparatory studies for wrc-23,'' \url{https://www.itu.int/en/ITU-R/study-groups/rcpm/Pages/wrc-23-studies.aspx}, 2023,
\newblock Accessed: 2024-03-25.

\bibitem{davies2021approaches}
Jack~BP Davies and Jonathan Woodburn,
\newblock ``Approaches to and loci for regulation of large and mega satellite constellations,''
\newblock in {\em Legal Aspects Around Satellite Constellations: Volume 2}, pp. 47--81. Springer, 2021.

\bibitem{millwood2023urgent}
Scott Millwood,
\newblock {\em The Urgent Need for Regulation of Satellite Mega-constellations in Outer Space},
\newblock Springer Nature, 2023.

\bibitem{pritchard2023wrc}
Ruth Pritchard-Kelly,
\newblock ``{WRC-23} on the horizon: Large satellite constellations, {ITU} issues, and industry perspective,''
\newblock {\em Air and Space Law}, vol. 48, no. Special, 2023.

\bibitem{10209992}
Mahdis Jalali, Flor Ortiz, Eva Lagunas, Steven Kisseleff, Luis Emiliani, and Symeon Chatzinotas,
\newblock ``Joint power and tilt control in satellite constellation for {NGSO-GSO} interference mitigation,''
\newblock {\em IEEE Open Journal of Vehicular Technology}, vol. 4, pp. 545--557, 2023.

\bibitem{cottatellucci2006interference}
Laura Cottatellucci, Merouane Debbah, Gennaro Gallinaro, Ralf Mueller, Massimo Neri, and Rita Rinaldo,
\newblock ``Interference mitigation techniques for broadband satellite systems,''
\newblock in {\em 24th AIAA International Communications Satellite Systems Conference}, 2006, p. 5348.

\bibitem{10293839}
Almoatssimbillah Saifaldawla, Flor~G. Ortiz-Gomez, Eva Lagunas, Saed Daoud, and Symeon Chatzinotas,
\newblock ``{NGSO-To-GSO} satellite interference detection based on autoencoder,''
\newblock in {\em 2023 IEEE 34th Annual International Symposium on Personal, Indoor and Mobile Radio Communications (PIMRC)}, 2023, pp. 1--7.

\end{thebibliography}
\section{Biography Section}
\vspace{11pt}

\vspace{11pt}
\begin{IEEEbiographynophoto}{Flor G. Ortiz} received her B.S. degree in telecommunications engineering and her M.S. degree in electrical engineering-telecommunications from the Universidad Nacional Aut\'onoma de M\'exico (UNAM), Mexico City, Mexico, in 2015 and 2016, respectively. Flor obtained her Ph.D. degree in Telecommunication Engineering (September 2021) at the Universidad Politecnica de Madrid (UPM), Madrid, Spain. During her PhD, she performed a research period at the University of Bologna, in Bologna, Italy. She started a close collaboration between UPM and the University of Bologna, opening a new research line for both groups on applying Machine Learning for radio resource management. She is joined as a Research Associate at the Interdisciplinary centre for Security, Reliability, and Trust (SnT) at University of Luxembourg. Her research interests are focused on implementing cutting-edge Machine Learning techniques including Continual Learning and and Neuromorphic Computing for operations in Satellite Communications systems.
\end{IEEEbiographynophoto}
\begin{IEEEbiographynophoto}{Eva Lagunas}
received the M.Sc. and Ph.D. degrees in telecommunications engineering from the Polytechnic University of Catalonia (UPC), Barcelona, Spain, in 2010 and 2014, respectively. She has held positions UPC, Centre Tecnologic de Telecomunicacions de Catalunya (CTTC) University of Pisa, Italy; the Center for Advanced Communications (CAC), Villanova University, PA, USA. In 2014, she joined the Interdisciplinary Centre for Security, Reliability and Trust (SnT), University of Luxembourg, where she currently holds a Research Scientist position. Her research interests include terrestrial and satellite system optimization, spectrum sharing, resource management and machine learning.
\end{IEEEbiographynophoto}
\begin{IEEEbiographynophoto}{Almoatssimbillah Saifaldawla}~ received his B.Sc. (Honors) degree in Electronics Engineering Technology from the University of Gezira (UofG), Sudan, in June 2018. He obtained his M.Eng. in Communication and Information Engineering from Chongqing University of Posts and Telecommunications (CQUPT), China, in June 2022. He is working toward a Ph.D. degree in Computer Science and Computer Engineering with the SIGCOM research group in the Interdisciplinary Centre for Security, Reliability and Trust (SnT), University of Luxembourg. His research interests include wireless communications, resource management, and machine learning for satellite communications. 
\end{IEEEbiographynophoto}
\begin{IEEEbiographynophoto}{Mahdis Jalali} received the B.Sc. and M.Sc. degrees from the Department of Electrical Engineering, Amirkabir University of Technology (Tehran Polytechnic), Tehran, Iran. She is currently working toward the Ph.D. degree with SIGCOM Group, SnT, University of Luxembourg. Her research interests include satellite communication, resource management, and machine learning approaches for wireless communication systems.
\end{IEEEbiographynophoto}
\begin{IEEEbiographynophoto}{Luis Emiliani} received the B.E.E. and M.Eng. degree from Universidad Pontificia Bolivariana, Medellin, Colombia, in 1999 and 2003, respectively. He is with SES S.A., Betzdorf, Luxembourg, in the Fleet Advancement Group responsible for setting and developing SES's communications satellite fleet strategy and leading the fleet planning process. His research interest focuses on the application of propagation techniques to solve practical sharing and system design scenarios. Mr. Emiliani is a Chartered Engineer (C.Eng.) with the U.K. Engineering council and Member of the Institution of Engineering and Technology.
\end{IEEEbiographynophoto}
\begin{IEEEbiographynophoto}{Symeon Chatzinotas} received the M.Eng. degree in telecommunications from the Aristotle University of Thessaloniki, Thessaloniki, Greece, in 2003, and the M.Sc. and Ph.D. degrees in electronic engineering from the University of Surrey, Surrey, U.K., in 2006 and 2009, respectively. He is currently a Full Professor/the Chief Scientist I and the Head of the SIGCOM Research Group, SnT, University of Luxembourg. In the past, he has been a Visiting Professor with the University of Parma, Italy, and he was involved in numerous research and development projects for the National Center for Scientific Research Demokritos, the Center of Research and Technology Hellas, and the Center of Communication Systems Research, University of Surrey. He has (co-)authored more than 400 technical papers in refereed international journals, conferences, and scientific books. He was a co-recipient of the 2014 IEEE Distinguished Contributions to Satellite Communications Award, the CROWNCOM 2015 Best Paper Award, and the 2018 EURASIC JWCN Best Paper Award. He is on the editorial board of IEEE Open Journal of Vehicular Technology and International Journal of Satellite Communications and Networking.
\end{IEEEbiographynophoto}

\vfill
\end{document}